# Multivariate Modeling of Daily REIT Volatility[*]

John Cotter, University College Dublin[†]
&
Simon Stevenson, Cass Business School, City University and
University College Dublin[‡]

[†] Centre for Financial Markets, Department of Banking & Finance, University College Dublin, Blackrock, County Dublin, Ireland, Tel: +353-1-7168900, Fax: +353-1-2835482,
E-Mail: john.cotter@ucd.ie

[‡] Real Estate Finance & Investment Group, Faculty of Finance, Cass Business School, City University, 106 Bunhill Row, London, EC1Y 8TZ, UK.
Centre for Real Estate Research, Smurfit School of Business, University College Dublin, Blackrock, County Dublin, Ireland, Tel: +353-1-7168848, Fax: +353-1-2835482,
E-Mail: simon.stevenson@ucd.ie

APRIL 25 2005

[*] The authors would like to acknowledge the comments from participants at the Second Annual Hong Kong-Singapore International Real Estate Research Symposium, in particular our discussant Shaun Bond, and to seminar participants at the Department of Finance, Goethe University, Frankfurt, Germany.

# Multivariate Modeling of Daily REIT Volatility

## Abstract

This paper examines volatility in REITs using a multivariate GARCH based model. The Multivariate VAR-GARCH technique documents the return and volatility linkages between REIT sub-sectors and also examines the influence of other US equity series. The motivation is for investors to incorporate time-varyng volatility and correlations in their portfolio selection. The results illustrate the differences in results when higher frequency daily data is tested in comparison to the monthly data that has been commonly used in the existing literature. The linkages both within the REIT sector and between REITs and related sectors such as value stocks are weaker than commonly found in monthly studies. The broad market would appear to be more influential in the daily case.

# Multivariate Modeling of Daily REIT Volatility

1. Introduction

This paper examines volatility in Equity REITs using a multivariate GARCH based model. The Multivariate VAR-GARCH technique documents the return and volatility linkages between REIT sub-sectors and also examines the influence of other US equity series. The motivation is for investors to incorporate time-varyng volatility and correlations in their portfolio selection. As Lee & Stevenson (2004) note, REITs to some extent provide a hybrid investment form, standing between equities and the fixed-income sector. In addition, the asset maintains strong links with the direct private real estate market[1]. These interlinkages provide the asset with unique characteristics. The literature that has examined the characteristics of REITs has largely concentrated upon the links with either the private market or mainstream equities. In addition, the literature has primarily been concerned with linkages in the return series' involved. This paper aims to extend the current literature by examining some of the key relationships in relation to volatility in the REIT sector. While a number of papers have examined interlinkages in volatility (Devaney, 2001, Stevenson, 2002 and Cotter & Stevenson, 2004), they have utilized univariate models whereas this paper extends the analysis of volatility dynamics by using a multivariate GARCH framework. This paper uses the BEKK process (Engle & Kroner, 1995) that isolates the bivariate volatility linkages for each REIT sub-sector. The advantage of the use of a multivariate framework is that not only does such an approach provide an analysis of volatility and the accompanying interlinkages between the assets concerned, but it also allows an estimation of the time-varying covariance's and correlations. This allows an investor to incorporate time-varying volatility and correlations in their portfolio selection and this paper explores their properties. Increased (decreased) correlation imply lower (higher) diversification effects impacting investor portfolio selection.

The rationale behind the examination of volatility and particularly the analysis of the issue on a daily basis is primarily motivated by the increased investor awareness in the asset. The development of the REIT sector in terms of a broader class of investor and the increase in trading may result in a changing relationship between the asset and the broader equity markets. In addition, factors such as the inclusion of REITs into mainstream benchmark indices may also result in general market sentiment being more influential in the determination of REIT returns than in the past. This potential changing environment will also have implications for issues such as risk measurement and management. As increases in daily trading occur this will in all likelihood alter the daily volatility patterns in REITs and due to the impact of volatility in key risk measures such as value-at-risk, a keen understanding of the underlying dynamics will increase in importance. If it is the case that volatility has increased in the REIT sector then not only will this make risk measurement and management more important, but in addition, it will make effective hedging more complex. This is particularly so in the case of an asset such as REITs that lack the availability of traded hedging instruments commonly found in other mainstream equity sectors.

The growth in investor awareness and interest in the REIT sector in recent years has encouraged a large literature to develop examining the characteristics of the asset. Much of this increased investor interest has taken place in recent years due to a combination of factors, including the strong performance of the asset, its relative performance against other equity classes and its inclusion in benchmark indices. As of the end of 2003 there were 171 registered REITs, with a market capitalization of $224bn. While much of the growth in the number of REITs took place in the late eighties and early nineties, the growing maturity and development of the sector can be observed by comparing current market capitalization's with those in the past. In 1990 the total REIT sector had a market capitalization of only $8.7bn, while even in 1995 the figure was still only $57bn.

If one compares the overall REIT sector to major other equity indices the strong relative performance of REITs becomes apparent. Against the S&P 500 and the Wilshire small and mid cap growth and value indices and the Wilshire micro-cap index, the NAREIT Overall Index outperforms on a risk-adjusted basis for each of the following periods; 1990-2003, 1995-2003 and 2000-2003[2]. The strong asset backing and the income properties of REITs, due in part to the requirement that a minimum of 90% of taxable income is distributed in dividends, have also been particularly attractive to fund managers in recent years due to the high volatility in capital markets and the technology crash that led to investors looking for more secure and income based investment vehicles. A further factor that also increased investor awareness and investment into the sector was the inclusion of REITs into the S&P's mainstream benchmark indices. Not only did this increase awareness but also led to increased investment from index-based funds. The combination of factors such as these, the limitations on REITs in relation to dividend payments and the strong relative performance of the sector in the aftermath of the collapse of the technology bubble have resulted in increased fund flows into the sector. Ling & Naranjo (2003) illustrate, using a Vector Autoregressive (VAR) framework, that in their post 1992 sub-sample capital flows into REITs do significantly influence subsequent returns in the REIT sector. Moreover, Ling & Naranjo (2004) report on the impact of the flow of funds into REITs and the subsequent impact upon REIT returns.

The remainder of the paper is laid out as follows. The following section reviews the relevant literature on REIT linkages. Section 3 details the data analyzed and reports basic summary statistics. In addition, the basic methodological approach is discussed and initial diagnostic results on the model are reported. Section 4 presents the main empirical analysis and results with regard to the multivariate GARCH model. The final section provides concluding comments.

**2. Literature Review**

The empirical evidence with regard to interlinking factors in REITs is similar to the broader literature to have assessed relationships in the wider capital markets. In particular, the evidence is similar to that reported in relation to linkages between international capital markets. In both strands of literature, the evidence with regard to linkages in returns has often been inconsistent. In contrast the relationships in volatility have been generally more consistent and often more intuitive in nature. A large number of papers have assessed the extent to which REITs move in line with mainstream stocks. While the majority of studies have broadly found evidence that the two markets are heavily linked, with REITs heavily dependent on movements in the mainstream market, there are a number of papers that have provided contrasting evidence. In particular, a large number of studies have illustrated that the early nineties saw a turning point in the price behavior of REITs, and in particular Equity REITs. To a large extent this shift was due to the reforms contained in the 1986 Tax Reform Act, which eliminated many of the tax based investment incentives of REITs. Prior to this legislation many REITs had been effectively established as tax shelters.

Early studies such as Li & Wang (1995), Liu et al. (1990) and Mei & Lee (1994) all reported strong relationships, often in terms of cointegration, between REITs and mainstream equities. Ling & Naranjo (1999) use multi-factor asset pricing techniques to examine whether there is any evidence of integration between direct real estate, REITs and common stocks. As with previous studies, REITs are found to be integrated with non-real estate equities, however, no such evidence is found in relation to the direct market, even when this data is adjusted for smoothing. Using a spectral analysis approach, Oppenheimer & Grissom (1998) find that equities have a dominant influence on REIT returns over the period from 1989 to 1995. Glascock et al. (2000) report that while REITs were segmented from the general equity market during the period 1972 to 1991, this was not the case in the following five years. For that period, 1992-1996, REITs were found to be integrated with common stocks.

However, while the above reported evidence is supportive of a strong relationship between REITs and equities, a number of studies have found contrasting evidence. Wilson & Okunev (1996) in their examination of not only REITs but also the indirect traded real estate markets in Australia and the UK, find in all three markets an absence of any cointegrating relationships. Okunev & Wilson (1996) use a non-linear integration test to examine the relationship between REITs and the S&P 500 Composite. The results show that while the two markets may be related in a non-linear fashion, the level of deviations between the two can be extensive, with the degree of mean reversion quite slow. For example, the authors find that the half-life of deviations is 30 months is some cases. In addition, Clayton & MacKinnon (2001) find that the sensitivity of REIT returns to the Stock Market declined significantly in the 1990s.

In contrast to this literature the empirical evidence with regard to relationships in volatility has been more intuitive. Stevenson (2002) examines volatility spillovers with regard to monthly REIT returns. The results with regard to the interlinkages with the mainstream capital market assets are intuitive, with small cap and value stocks being far more influential on REITs than the large cap S&P 500 Composite Index. While this is consistent with the literature to have highlighted the similarities between REITs and value stocks in the mid and small cap range, such as Chiang & Lee (2002), a recent paper by Cotter & Stevenson (2004) shows that these findings are sensitive to the frequency of data analyzed. While Stevenson (2002) uses the monthly NAREIT indices, Cotter & Stevenson (2004) use daily data. In their analysis of univariate relationships the large cap sectors are found to be more influential in terms of the dynamics of REIT volatility. The authors hypothesize that the more intuitive results found in Stevenson (2002) which illustrate more fundamental linkages are to some extent masked when analyzing higher frequency data, with general market sentiment coming more to the fore. A number of other recent papers have also analyzed aspects of REIT volatility, however, they have largely concentrated upon the determinants of volatility and not the interlinkages with other asset classes. Devaney (2001) uses a GARCH-M model on monthly REIT data, primarily to examine the relationship between REIT volatility and interest rates. Two recent working papers, Winniford (2003) and Najand & Lin (2004) both provide further evidence concerning the dynamics of daily volatility in the REIT sector. More generally speaking a large literature has provided evidence of volatility spillovers. Studies such as Hamao et al. (1990), Lin et al. (1994) and Bekeart & Harvey (1997) all document the presence of spillovers in international capital markets. Further studies have also examined the foreign exchange markets, Melvin & Melvin (2003) and Huang & Yang (2002).

As well as the relationships between REITs and other equity sectors and the fixed income and cash markets, this paper also analyses the situation with REITs and the linkages between equity, mortgage and hybrid REITs. In terms of returns, a number of studies have found evidence of strong interlinkages between REIT sectors. He (1998), for example, finds evidence to support the notion that a causal relationship exists from Equity REITs to Mortgage REITs in the USA, with further evidence finding that the two sectors are cointegrated. Lee & Chiang (2004) also find evidence of commonalities between the equity and mortgage sectors. The authors use the variance decomposition approach proposed by Seck (1996) and variance ratio tests, with the results supporting the hypothesis that the two sectors are substitutable. Where studies do differ is often in relation to the apparent structural shift that occurred in the early nineties. Glascock et al. (2000) report that while the sectors were substitutable prior to 1992, with evidence of cointegration between the two sectors and common driving forces, this affect is not evident in the post-1992 environment. This finding is consistent with the other results from Glascock et al. (2000), previously discussed, concerning the changing relationship between Equity REITs and the general equity market, with a structural break occurring in the price behavior of the asset in the early nineties.

## 3. Data

The empirical tests conducted in this paper utilize the NAREIT Daily indices, which currently extend back historically to January 1999. The dataset used comprises of data for 4½ years, from January 1999 through to June 2003. As a key element of the analysis is the examination of the interlinkages between the different REIT sectors, the overall REIT index is not examined in the main empirical analysis. The investigation rather is focused on the three main sub-sectors, namely the Equity, Mortgage and Hybrid sectors. The classification of these sectors is determined as follows. Equity REITs are required to have 75% of their invested assets in equity real estate investments, while Mortgage REITs are required to hold a minimum similar percentage in debt secured on real estate. Hybrid REITs are those trusts that do not satisfy the above two criteria. The exclusion of the All REIT Index is also based on the fact that as of the end of 2003, out of a total of 171 REITs, 144 were classified as Equity REITs, with this sector accounting for 91% of the total REIT market capitalization. Therefore, the Equity and All REIT Indices are effectively one and the same.

The equity indices analyzed include the S&P 500 Composite, S&P mid-cap Value Index, S&P mid-cap Growth Index and the NASDAQ Composite. The S&P 500 is used as a proxy for the overall market and in particularly, the large cap sector, while the NASDAQ proxies 'new economy' firms. The analysis of the value sector is motivated by the similarities between value stocks and REITs and in particular the high level of asset backing in both sectors. In addition, as few REITs currently can be classified as large cap, the use of a mid-cap value index is also deliberate to avoid any spurious results due to differences in the market capitalization of the constituent stocks in the two indices. A large number of studies have illustrated the linkages between value stocks and REITs, including Chiang & Lee (2002), who illustrate using Style Analysis that Equity REITs can be classified as a combination of value stocks and treasury bills. In addition, Stevenson (2002) in his analysis of volatility spillovers in REITs using monthly data finds strong evidence that value stocks provide more significant influence on REIT volatility than the large cap or growth sectors. The paper specifically does not include bond sectors in the analysis. This is due to two primary reasons. Firstly, the motivation of the paper is primarily concerned with the interlinkages between REITs and other equity sectors and secondly, Cotter & Stevenson (2004) found little evidence of interlinkages in daily volatility in their univariate analysis. In order to avoid spurious conclusions due to mis-specification of the data, stationarity tests were conducted with the Augmented Dickey Fuller (ADF) unit root tests on all of the series analyzed. The findings show that the price data does not accept the hypothesis of stationarity, however, once first differenced, the return series are stationary.[3]

{Insert Table 1}

Some descriptive statistics of the respective series are outlined in Table 1 detailing the first four moments of each series, and the correlation matrix between the series. For comparative purposes the S&P 500 Composite is also included. The average returns illustrate that with the exception of Hybrid REITs, the sector provides positive average daily returns, while each of the three REIT series' outperforms the S&P 500 index. Given the time period examined, and that it contains the technology crash, the poor relative performance of the general equity market is not surprising. In addition, REITs have provided one of the best return profiles in recent years due to their asset backing and high income potential. Mortgage REITs surprisingly have higher unconditional volatility than other indexes with daily volatility of 1.6%. The usual features of excess skewness and kurtosis are exhibited for all series, being most pronounced for the Mortgage series. The correlation coefficients show, as expected, that Mortgage REITs consistently have the lowest coefficients with the other asset classes, while the highest correlations reported are those between Equity REITs and both the S&P and the Hybrid sector. It is however noticeable that the reported correlation between the Equity REITs and the general equity market is lower than perhaps expected. This could be due to the use of daily rather than lower frequency data and also the specific time period examined.

{Insert Figure 1}

Time series plots of daily returns are given in Figure 1. The pattern for Equity and Hybrid REITs follow a similar pattern, with returns constrained to a similar magnitude. All returns are time varying with volatility clusters. For instance, the highest level of volatility occurred towards the end of 2002 for Equity REITs whereas Mortgage REITs incurred very high volatility during middle of 1999 with returns in excess of 20% and –10%[4]. Dependence in the returns and volatility series is examined by the autocorrelation function (ACF) over 30 lags and is displayed in Figures 2 and 3. Squared returns are used to detail characteristics of the volatility series. Similar findings in line with financial time series in general are identified with low persistence in returns being contrasted by relatively strong persistence in the volatility series. This finding is associated with the volatility clustering property of financial time series. The strong serial correlation of volatility indicates the existence of ARCH effects and validities the application of GARCH related processes. While there is significant dependency of returns in the first lag due to non-synchronous trading, there is a general lack of significant autocorrelation in all returns series. Persistence is strongest for Equity REITs returns. Also the persistence of this series volatility follows a similar pattern to financial time series in general with strong significant dependence decreasing slowly and remaining over the first ten lags and thereafter being insignificant. Persistence in volatility is weakest for Mortgage REITs although it does record the largest single autocorrelation estimate.

{Insert Figures 2 & 3}

The empirical analysis is undertaken using a multivariate GARCH (Generalized Autoregressive Conditional Heteroscedasticity) framework. GARCH models allow the simultaneous modeling of both the first and second moments of the return series' and provide a more efficient means of modeling time-series'. The use of such models allows an examination of the interlinkages between the different assets in terms of their second moment. A basic univariate GARCH (1,1) model can be displayed as follows:

$$R_t = bZ_t + \varepsilon_t \tag{1}$$

$$\varepsilon_t | I_{t-1} \sim N(0, H_t) \tag{2}$$

$$H_t = \alpha_0 + \sum \alpha_i \varepsilon_{t-i}^2 + \sum \beta_j H_{t-j}^2 \tag{3}$$

Where the mean is described by a first order VAR, and univariate volatility follows a GARCH process. The basic univariate GARCH model can also be extended in a variety of ways such as the EGARCH (Exponential GARCH) and GARCH-in-mean specifications. The EGARCH specification examines what is referred to as the leverage effect. This is concerned with whether volatility is affected asymmetrically by the impact of positive and negative news. The first paper to note this asymmetry was Black (1976) who reported that volatility increased (decreased) as a result of periods of bad (good) news announcements. In essence, asset returns and future volatility are negatively related. Cotter & Stevenson (2004) examine asymmetry in REIT volatility using a univariate EGARCH model. However, the authors find no evidence of a leverage effect. While it would be possible to incorporate the analysis of the leverage effect into the multivariate framework used in this paper due to these previous findings this extension is not undertaken. Multivariate GARCH models have an advantage over univariate versions in that they also allow modeling of the time-varying covariances/correlations. This allows an examination of how the relationship between the assets concerned alters throughout the time series utilized. The multivariate model used in this paper is the BEKK specification proposed by Engle & Kroner (1995). The model can be written as follows:

$$H_{i,t} = C_0^{*'}C_0^{*} + A_{11}^{*'}H_{i,t-1}A_{11}^{*} + B_{11}^{*'}\varepsilon_{i,t-1}\varepsilon_{i,t-1}^{'}B_{11}^{*} \tag{4}$$

Detailing the bivariate volatility linkages for $H_{i,t}$ is the conditional variance covariance matrix at $t$. Each matrix, $C$, $A$ and $B$ is 2 x 2 and $C$ is restricted to be upper triangular. Hence we have 11 free parameters in the model and $H_{i,t}$ positive definite is assured. It is because of the assurance that the H matrix is positive definite that the BEKK specification is used in preference to alternative multivariate models such as the VECH specification. Return linkages between the REIT series are ascertained by the VAR parameters. Volatility spillover effects are ascertained from the GARCH estimates. The BEKK model implies that only the magnitude of past return innovations is important in determining current time-varying variances and covariance's irrespective of sign. The BEKK model has two important advantages over traditional multivariate approaches. First, it generates a time-varying correlation structure, as one would expect from the linkages between different assets over time in contrast to the constant conditional models. Second, it is easy to estimate with parsimonious modeling of just 11 parameters describing the full set of bivariate interactions between the series where convergence is easily achieved by this framework. Third, the BEKK model guarantees a positive variance-covariance matrix.[5]

Some pre and post diagnostics determine the suitability of the BEKK model to the respective returns series and are reported in Table 2. The results are supportive of the GARCH modeling. Prior to running the GARCH model the existence of heteroscedasticity is tested for on the return series using two tests. First, the Engle (1982) LM test for ARCH of order *p* tests the null of zero slopes in the regression:

$$y_{i,t}^2 = \phi_0 + \sum_{m=1}^{p} \phi_m y_{i,t-m}^2 + u_{i,t} \tag{5}$$

The test is performed as $T \cdot R^2$ where T represents the sample size. Second, the Ljung-Box test is utilized to determine the degree of dependence in the second moment of the series. Whilst there is no evidence of excessive dependence in the first moment of the returns series themselves with the exception of the Hybrid REIT and S&P500 series, the null of no serial correlation is rejected for all series indicating the suitability of modeling the returns with a GARCH specification. Furthermore ARCH effects are indicated for all series using the LM test.

{Insert Table 2}

After fitting the GARCH specification, further diagnostics are computed. These show that the BEKK model is generally well specified with either the removal or a dramatic reduction of serial dependence in the residual series. The Engle (1982) test for ARCH effects indicates the removal of significant dependence for the squared residual series for all series. Finally some summary statistics of the standardized returns series suggest near, but not quite, normality after fitting the BEKK model with a large reduction in the excess skewness and excess kurtosis statistics. The kurtosis values however, suggest that the GARCH model is not completely able to remove the fat-tailed feature of the returns data.

**4. Empirical Analysis**

The main empirical analysis initially examines the relationships between the different REIT sectors, and is then extended to analyze bivariate GARCH models between each of the REIT sectors and the equity sectors examined[6]. Prior to fitting the model, the returns series are put through an MA filter. While the filter in itself has little impact on the VAR-GARCH results its coefficient is significant for the following series: Equity and Mortgage REITs (0.084), Equity and Hybrid REITs (0.174), Mortgage and Hybrid REITs (0.084), Equity REITs and S&P 500 (0.173), and Mortgage REITs and S&P 500 (0.085). The VAR-GARCH model is also fitted assuming unconditional normality and allowing for fat-tails, an unconditional t-distribution. There are no discernable differences between the two models and the former values only are reported. The bivariate return and volatility relationships between the different REIT sectors are outlined in Table 3. Panel A of the table reports the results from the conditional mean equation, while the conditional variance equation findings are reported in Panel B. The impact of an asset's own market effects are represented by 11 for asset 1, and 22 for the second asset class. Cross-market effects are given by 21 and 12. In the conditional variance equation the *a* coefficients represent ARCH effects, while the *b* coefficients are GARCH effects. As would be expected in both the return and volatility equations, past own series returns impact both current returns and volatility. In each case the appropriate coefficients are significant, supporting the previously reported findings with regard to autocorrelation in the returns. The findings are supportive of the hypothesis of volatility clustering, with an autocorrelated relationship present in the second moment of the distribution. The VAR also examines the direction and magnitude of the return linkages. There is some evidence of significant positive interlinkages, and in an anticipated direction. Significant findings are reported with respect to both Equity and Mortgage REITs significantly impacting upon the smaller Hybrid sector. As would be expected this is not a reciprocal relationship, with no evidence of Hybrid REITs having a significant impact upon the two larger sectors. Likewise, the relationship between Equity and Mortgage REITs is not significant, however, due to the differences in the composition of the REITs concerned, this is not surprising. The results highlight the fact that the market appears to effectively distinguish between different REIT sub-sectors and acknowledges the fundamental differences in the composition of Equity and Mortgage REITs. The results, although modeled in a different context, also support the findings of studies such as He (1998) who found evidence of causal relationships in REIT returns. Turning to the volatility modeling the main diagonal elements of the variance covariance matrix are typical of a GARCH process. Autoregressive and time dependent volatility effects exist for each series according to the a11, a22, b11, and b22 parameters.

{Insert Table 3}

It is interesting that the volatility results are less conclusive than those reported with the return series'. The results indicate that there are relatively weak spillover effects between the REIT series from the examination of the off-diagonal elements of the covariance matrix. Furthermore the cases that do provide significant findings are not intuitive. No significant coefficients are reported in the direction of Hybrid REITs, nor do the Equity and Mortgage Sectors note any relationships. Rather, the primary sources of spillover effects are Hybrid REITs, in the direction of the Equity Sector. Given that Equity REITs comprise over 90% of the total REIT sector, and are generally the largest and most heavily traded stocks in the sector, this result is somewhat perplexing. Previous studies such as Stevenson (2002) have in some cases found a spillover effect from Hybrid REITs to Mortgage REITs, however, this could be explained by the fact that the Hybrid Sector captures some of the effects of the Equity sector due to the mixed nature of the REITs concerned. The results reported here however cannot be explained in a similar fashion. It should be noted that the number of REITs classified as Hybrid's has reduced dramatically in recent years, and as of the end of 2003 only seven REITs were thus classified. This is out of a total of 171 REITs, meaning that Hybrid's comprise 4.09% of the total number of REITs. In comparison, in 1993, 22 REITs were classified as Hybrid's, making up 11.64% of the total. In addition, the average market capitalization of Hybrid's is just over half that of Equity REITs, $746m in comparison to $1,422m. Therefore, given the small population, the results with regard to the Hybrid sector should be approached with some caution.

More generally speaking the results are in marked comparison to Stevenson (2002) and his analysis using monthly REIT data. Stevenson (2002) found very few insignificant results, with the majority of relationships showing significant volatility spillovers. These findings were generally consistent whether lagged returns were incorporated into the analysis or not and whether a GARCH or EGARCH specification was used. Equity REITs were found to significantly influence volatility in the Mortgage and Hybrid sectors in all but one of the scenario's examined, while Hybrid REITs provided significant spillover effects to Mortgage REITs in each case, with a significant bi-directional relationship reported in one case. The contrasting findings reported in the current paper are however supported by the univariate analysis contained in Cotter & Stevenson (2004). This paper also found a distinct lack of evidence supporting the hypothesis of volatility spillovers in daily data. The results found in both the current paper and in Cotter & Stevenson (2004) would indicate that the use of different data frequency could lead to very contrasting empirical findings. Monthly data would appear to allow more time for the more substantial and intuitive relationships to come to the fore. It is possible that the use of the higher frequency data masks more of these fundamental relationships, with general market sentiment coming more to the fore. The use of daily data may also help to explain the results regarding Hybrid REITs, due to their small size and thin trading. This is an issue that will be explored further in the following empirical analysis that examines the relationship between REITs and other equity sectors.

As well as examining relationships between REIT series, the dynamics between some common equity series including S&P500, S&P Value, S&P Growth and NASDAQ indexes with the REIT series' are also analyzed. Bivariate return and volatility linkages are outlined in Table 4. As expected the influence of own strong past autoregressive and volatility effects for the equity series is supported, however, as with the within REIT analysis, the relationships observed between sectors are generally weak. In addition, the relationships reported are not consistent across the three REIT sectors. One of the more intuitive findings is that Equity REITs is the sub-sector most influenced by the general equity sectors. In terms of returns, significant influences are observed from both the S&P500 Composite and the S&P Value series, with weaker findings reported for the S&P Growth Index and the NASDAQ Composite. These results are highly intuitive. Firstly, the impact of the general market is confirmed, while in terms of the more specialized sectors, the similarities between REITs and value stocks is also borne out. This was a key finding of Stevenson (2002), who highlighted the similarities between REITs and value stocks in relation to volatility and to the broader literature detailed earlier in the paper concerning linkages and similarities in return structures. Both sectors will generally have relatively high levels of asset backing and income flows in relation to their stock price. In addition, the performance of the stock market during the period examined provides further rationale behind these findings. The period examined includes the technology crash, which would obviously impact primarily upon the growth sector and the NASDAQ. Given that fund managers then switched large allocations into value sectors, including REITs, it is not surprising that diverging results are reported with respect to growth stocks and the NASDAQ, and strong links are noted with the value sector.

{Insert Table 4}

In contrast the findings for the Mortgage and Hybrid sectors provide only one significant finding. This is in the case of the NASDAQ influencing Hybrid REITs. In each other case there is an insignificant coefficient, indicating a lack of interlinkages. Again, this makes intuitive sense in that Equity REITs are that type of REIT most similar to the general stock market and that submarket most likely to be influenced by general market sentiment and trends. In addition, the sector is, as already noted, the largest both in terms of the number of securities and both aggregate and average market capitalization. The higher levels of trading thus observed in the equity sector would mean that there is an inclination for that sector to be more influenced by general market sentiment. The one significant finding reported is actually negative, with the NASDAQ having a negative impact upon the Hybrid sector. Again, this is probably due to the market conditions during the period under study and the divergence in performance.

The results with regard to volatility interactions are more mixed and are less intuitive when viewed initially. However, they also generally support the findings reported by Cotter & Stevenson (2004) in their univariate analysis. No significant coefficients are reported with respect to Hybrid REITs, while in relation to the Equity and Mortgage sectors two significant findings are reported in each case. In both cases the overall S&P 500 provides the cause of significant volatility spillovers to the REIT sector. Whilst with the Mortgage sector a further significant coefficient is reported with respect to value stocks, in the case of the Equity sector the second significant finding is with regard to the NASDAQ. These results do need some form of explanation. In terms of the mortgage sector, while initially this appears intuitive, it needs careful examination. Given the different characteristics of the Mortgage sector in comparison to Equity REITs, the link with value stocks is not actually as strong or as intuitive. However, these results can perhaps be explained due to the diverging performance in the market during this period. The poor performance of the growth side of the market meant that funds flowed into the value side of the market, but more importantly, also the fixed income market. Therefore, what we are perhaps witnessing here is an indirect impact in that value stocks and the bond and money markets were behaving in a similar fashion volatility wise due to a flow of funds effect. The results in relation to the Equity sector are not intuitive but again support the hypothesis proposed in Cotter & Stevenson (2004), who also found similar results. This is the argument, which was initially discussed during the within REIT analysis, that the use of the higher frequency data may mask some of the more intuitive relationships at play. These then only come to the fore when lower frequency data is tested, as in the case of the Stevenson (2002) study that used monthly returns. If this were the case then it would appear that in terms of volatility, on a daily basis general market sentiment, represented in this case both by the S&P500 and the NASDAQ is far more influential than submarket volatility, even if these relationships may appear to be more intuitive. As discussed earlier, the larger size and more active trading in Equity REITs also potentially plays a role in this case.

The second section of the analysis examines the implied correlations derived from the multivariate BEKK model. Summary statistics for daily REIT conditional correlations are outlined in Table 5. Investors need to incorporate the fact that increased (decreased) correlation leads to a reduced (increased) diversification effects. It can be seen that the correlations have a high variance and deviate substantially. The deviation can be illustrated by comparing the minimum and maximum coefficients reported in Table 5. While the low coefficients reported in cases between the mortgage sector and the equity and hybrid sectors is to some extent expected, the range in the coefficients is extensive. In the case of the equity and mortgage sectors the correlations range from a minimum of –0.512 to a maximum of 0.909. While the maximum figures are similar across all three cases, the minimum figures are all negative even in the equity-hybrid case. While the use of daily data means that these correlations are not necessarily going to impact upon portfolio decision-making, as would perhaps be the case with monthly data, it does have implications in terms of risk measures, such as VaR, and also hedging decisions in deciding the time-varying optimal hedge ratio. Generally speaking the relationships involving the mortgage sector are the weakest, and extreme negative coefficients are recorded in some cases. However, despite the presence of negative coefficients they do largely remain positive, with the majority displaying positive but weak coefficients. Indeed, overall 75% of the reported coefficients are of a positive sign. The distributional properties of all linkages are characterized as being non-normal, with excess kurtosis and skewness, with these findings strongest for the equity sector. The findings reported in Table 5 highlight the advantages behind the use of a multivariate GARCH framework in the analysis of the time-varying correlation structure of the bivariate relationships.

{Insert Table 5}

Daily conditional correlation plots between REIT series extracted from the GARCH model are given in Figures 4 and 5. Linkages between both Equity and Mortgage REITs and the hybrid series have trended upwards over time reducing the potential for diversification across sectors. Again, this could be due to the decrease in the number of Hybrid REITs and the small numbers contained in this sample, which makes the results more susceptible to idiosyncratic risk. With regard to the mortgage sector it is noticeable that the correlations tend to be lower than the relationships observed between equity and hybrid REITs. Given the different characteristics of Mortgage REITs, this is not surprising.

{Insert Figures 4 & 5}

Daily conditional correlation plots between REIT and the S&P given in Figure 5[7]. The results are generally consistent. In relation to the equity sector, all correlation coefficients are positive peaking at approximately 0.7 with an increasing trend in more recent times. Similarly the relationships between the S&P and Mortgage REITs has increased over time but negative linkages occurred at the start of the sample. This description is generally true for the relationships incorporating Hybrid REITs but the correlations are not as strong in magnitude. This is possibly due to a number of factors. Firstly, the time period examined. The data used in this study goes from 1999 through 2003. This period saw generally poor performance from much of the general equity market and particularly the growth and technology sectors. Because of this fund managers looked to switch allocations to other capital market sectors. Due to their already strong performance, the increased flow of funds went into REITs, which further strengthened their performance. A second possible reason is the inclusion in 2001 of REITs in the S&P general market indices. Prior to this time REITs had been excluded from mainstream equity market indices. Their inclusion would have further raised awareness about the asset as well as attract increased flow of funds from index-based funds[8]. The implication of the upward trend in the correlations observed in relation to REITs and other equity classes has a number of implications. While it is too early to say whether this shift is a permanent structural change, given the potential causes it would not be unreasonable to assume that some form of structural break has occurred. Two primary implications would result from such a shift. Firstly, it would impact upon the diversification opportunities that REITs are perceived to provide within a capital market portfolio. This would not only be affected by an upward trend in the correlation coefficients but also in the general volatility of the sector. Secondly, these effects would imply that in some manner the sector is behaving in a more similar fashion relative to general stocks than in the past and in particularly is perhaps being driven by general market sentiment to a greater extent than in the past. This would also have implications in terms of the relationship between REITs, and especially the equity sector, with the underlying private real estate market.

## 5. Conclusion

This paper has examined the relationships between different REIT sectors in the context of both univariate and multivariate models. The results highlight the linkages between the different sub-markets and also the S&P 500 and other mainstream stock indices. The use of daily data highlights differences in the results in comparison to previous studies. The use of higher frequency data would appear to mask the intuitive findings reported in relation to the links both within the REIT sector and between REITs and the broader equity markets. It would appear that more general equity market sentiment plays a stronger role on a daily basis, with the linkages between REITs and related sectors such as value stocks less evident. Increased volatility puts pressure on investors to try and diversify some of this risk. The finding of increased correlations however makes this a difficult proposition for investors in their portfolio selection. The finding of rising correlations is an area that requires further examination, and in particular whether the trends noted are unique to the time period examined or whether they are more structural in nature.

**References**

Barkham, R. & Geltner, D. (1995). Price Discovery in American and British Property Markets, *Real Estate Economics*, **23**, 21-44.

Bekeart, G. & Harvey, C. (1997). Emerging Equity Market Volatility, *Journal of Financial and Quantitative Analysis*, **25**, 203-15.

Bollerslev, T. & Woodridge, J. (1992). Quasi-Maximum Likelihood Estimation and Inference in Models with Time Varying Covariance's, *Econometric Review*, 11, 143-172.

Black, F. (1976) Studies in Stock Price Volatility Changes, *Proceedings of the 1976 Business Meeting of the Business and Economics Statistics Section, American Statistical Association*, 177-181.

Bond, S. & Hwang, S. (2003) A Measure of Fundamental Volatility in the Commercial Property Market, *Real Estate Economics*, **31**, 577-600.

Chiang, K. & Lee, M. (2002). REITs in the Decentralised Investment Industry, *Journal of Property Investment & Finance*, **20**, 496-512.

Clayton, J and MacKinnon, G. (2001) The Time-Varying Nature of the Link between REIT, Real Estate and Financial Returns, *Journal of Real Estate Portfolio Management*, **7**, 43-54.

Cotter, J. & Stevenson, S. (2004). *Uncovering Volatility Dynamics in Daily REIT Returns*, Working Paper, Centre for Real Estate Research, Smurfit School of Business, University College Dublin.

Devaney, M. (2001) Time Varying Risk Premia for Real Estate Investment Trusts: A GARCH-M Model, *Quarterly Review of Economics & Finance*, **41**, 335-346.

Engle, R.F. (1982). Autoregressive Conditional Heteroscedasticity with Estimates of the Variance of UK Inflation, *Econometrica*, **50**, 987-1008.

Engle, R.F. & Kroner, K.F. (1995). Multivariate Simultaneous Generalised GARCH, *Econometric Theory*, **11**, 122-150.

Glascock, J., Lu, C. & So, R. (2000). Further Evidence on the Integration of REIT, Bond and Stock Returns, *Journal of Real Estate Finance & Economics*, **20**, 177-194.

Hamao,Y., Masulis, R.W., & Ng, V. (1990). Correlations in Price Changes and Volatility across International Stock Markets, *Review of Financial Studies*, **3**, 281-308.

He, L. (1998). Cointegration and Price Discovery Between Equity and Mortgage REITs, *Journal of Real Estate Research*, **16**, 327-338.

Huang, B., Yang, C. (2002). Volatility of Changes in G-5 Exchange Rates and Its Market Transmission Mechanism, *International Journal of Finance and Economics*, 37-50.

Lee, M.L. & Chiang, K. (2004). Substitutability Between Equity REITs and Mortgage REITs, *Journal of Real Estate Research*, **26**, 96-113.

Lee, S. & Stevenson, S. (2005). The Case for REITs in the Mixed-Asset Portfolio in the Short and Long Run, *Journal of Real Estate Portfolio Management*, **11**, 55-80.

Li, Y. & Wang, K. (1995). The Predictability of REIT Returns and Market Segmentation, *Journal of Real Estate Research*, **10**, 471-482.

Lin, W.L., Engle, R. & Ito, T. (1994). Do Bulls and Bears Move across Borders/ International Transmission of Stock Returns and Volatility, *Review of Financial Studies*, **3**, 507-538.

Ling, D. & Naranjo, A. (1999). The Integration of Commercial Real Estate Markets and Stock Markets, *Real Estate Economics*, **27**, 483-515.

Ling, D. & Naranjo, A. (2003). The Dynamics of REIT Capital Flows and Returns, *Real Estate Economics*, **31**, 405-434.

Ling, D. & Naranjo, A. (2004). *Dedicated and Non-Dedicated Mutual Fund Flows and REIT Performance*, Paper presented at the American Real Estate & Urban Economics Association Annual Conference (ASSA Meetings).

Liu, C.H., Hartzell, D.J., Greig, W. & Grissom, T. (1990). The Integration of the Real Estate Market and the Stock Market: Some Preliminary Evidence, *Journal of Real Estate Finance & Economics*, **3**, 261-282.

Mei, J. & Lee, A. (1994). Is there a Real Estate Risk Premium ?, *Journal of Real Estate Finance & Economics*, **9**, 113-126.

Melvin, M. & Melvin, B. (2003). The Global Transmission of Volatility in the Foreign Exchange Market, *Review of Economics and Statistics*, **85**, 670-79.

Najand, M. & Lin, C. (2004). *Time Varying Risk Premium for Equity REITs: Evidence from Daily Data*, Working Paper, Old Dominion University.

Okunev, J. & Wilson, P. (1997). Using Nonlinear Tests to Examine Integration Between Real Estate and Stock Markets, *Real Estate Economics*, **25**, 487-504.

Oppenheimer, P. and Grissom, T.V. (1998) Frequency Space Correlation between REITs and Capital Market Indices, *Journal of Real Estate Research*, **16**, 291-309

Seck, D. (1996). The Substitutability of Real Estate Assets, *Real Estate Economics*, **24**, 75-95.

Stevenson, S. (2002). An Examination of Volatility Spillovers in REIT Returns, *Journal of Real Estate Portfolio Management*, **8**, 229-238.

Wilson, P. & Okunev, J. (1996). Evidence of Segmentation in Domestic and International Property Markets, *Journal of Property Finance*, **7**, 78-97.

Winniford, M. (2003). *Real Estate Investment Trusts and Seasonal Volatility: A Periodic GARCH Model*, Working Paper, Duke University.

**Tables**

### Table 1: Descriptive Statistics of Daily Returns Series

| | Panel A: Moments | | | |
|---|---|---|---|---|
| | **Mean** | **Variance** | **Skewness** | **Kurtosis** |
| S&P500 | -0.037 | 1.949 | 0.156 | 4.080 |
| Equity REITs | 0.004 | 0.543 | 0.345 | 7.545 |
| Mortgage REITs | 0.011 | 2.572 | 1.854 | 65.850 |
| Hybrid REITs | -0.009 | 1.047 | 0.009 | 5.044 |
| | Panel B: Correlation Matrix | | | |
| | **S&P500** | **Equity REITs** | **Mortgage REITs** | **Hybrid REITs** |
| S&P500 | 1.000 | 0.471 | 0.218 | 0.345 |
| Equity REITs | 0.471 | 1.000 | 0.287 | 0.494 |
| Mortgage REITs | 0.218 | 0.287 | 1.000 | 0.250 |
| Hybrid REITs | 0.345 | 0.494 | 0.250 | 1.000 |

Notes: The first two moments are expressed in percentage form. The skewness and kurtosis statistics have a value of 0 for a normal distribution. The covariance and correlation matrices give a preliminary indication of the linkages between the indexes.

**Figure 1: Time Series Plots for Daily REIT Returns**

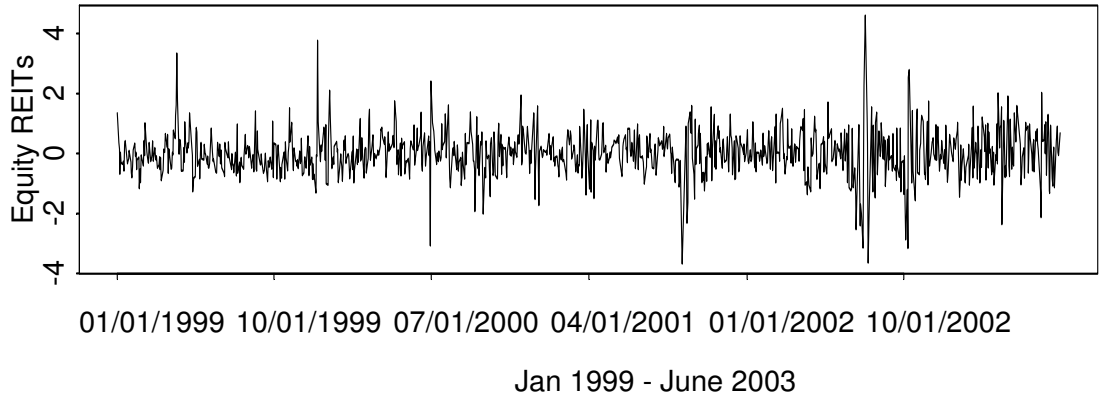

Jan 1999 - June 2003

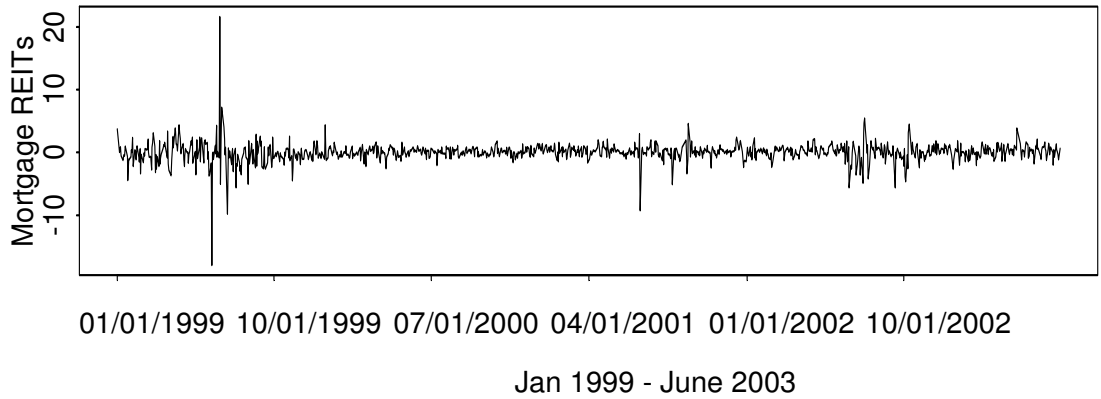

Jan 1999 - June 2003

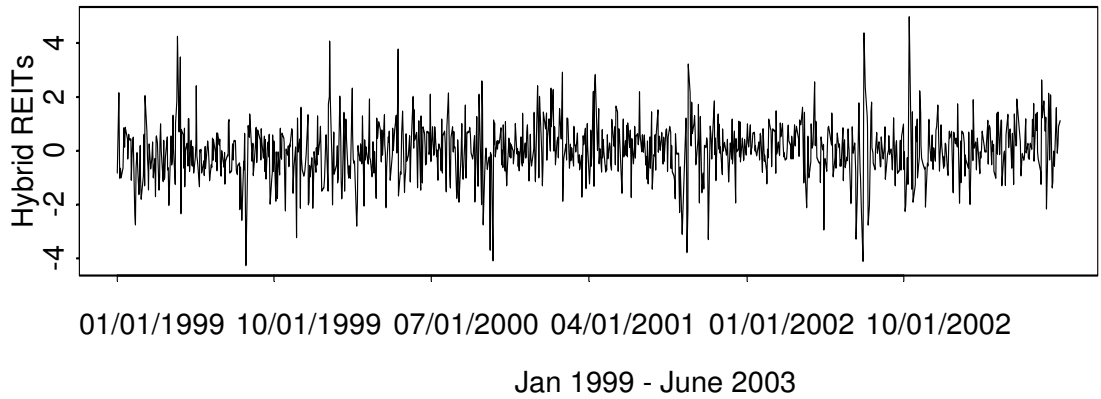

Jan 1999 - June 2003

**Figure 2: Autocorrelation Plots for Daily REIT Returns**

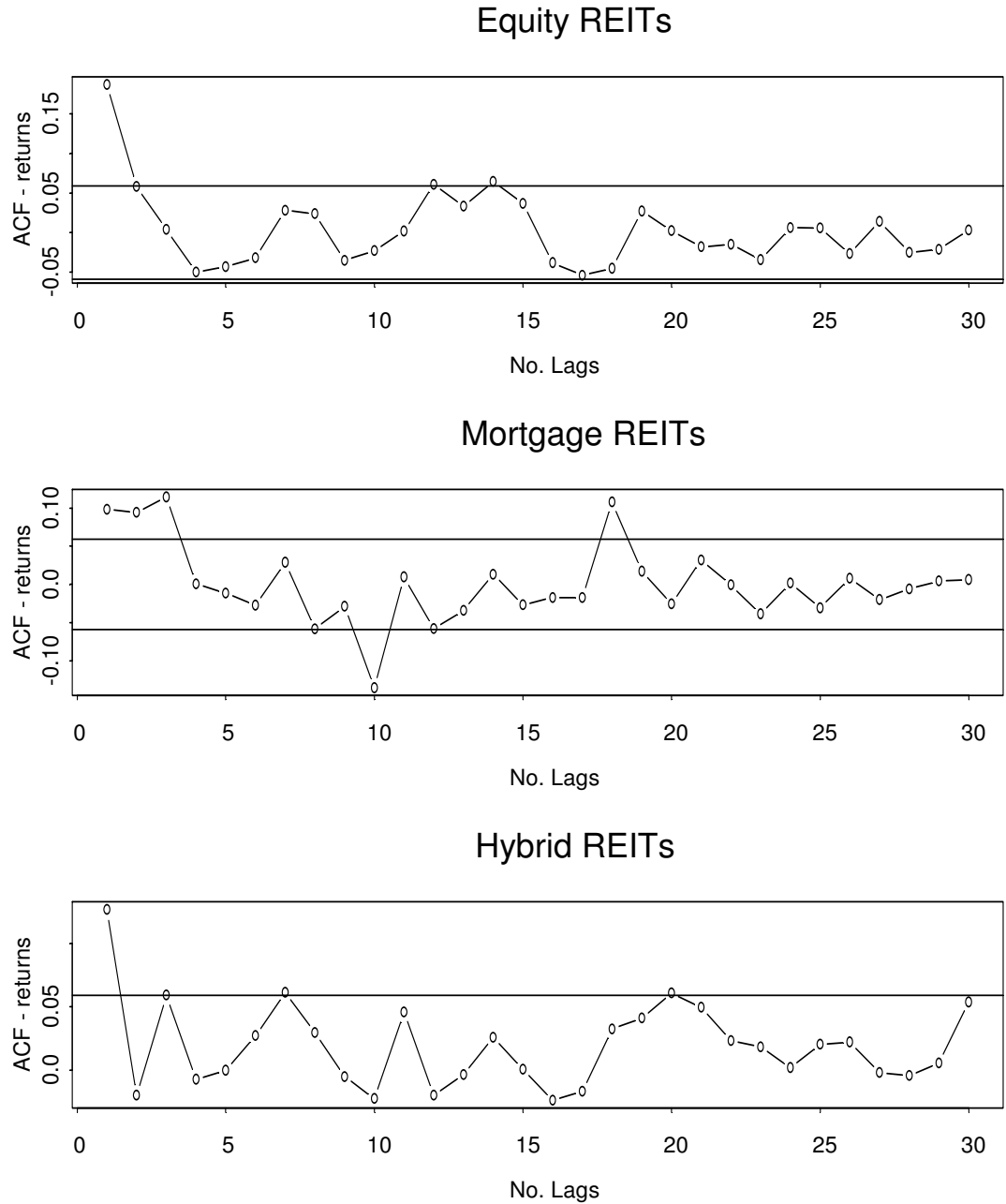

Notes: All plots include confidence bands measured by ± 1.96/√T so significance occurs at ± 0.059 and these are imposed where appropriate.

**Figure 3: Autocorrelation Plots for Daily REIT Volatility**

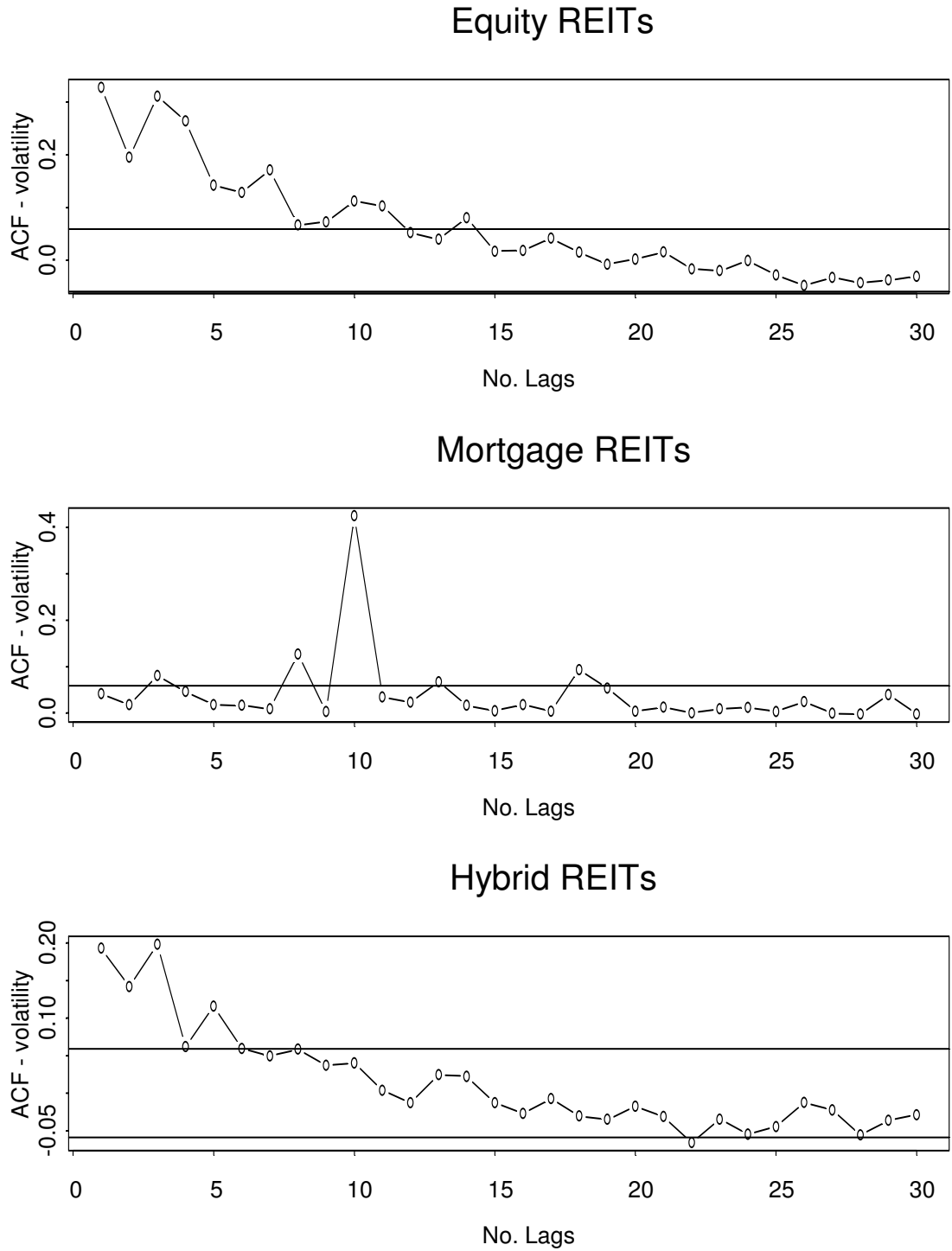

Notes: All plots include confidence bands measured by ± 1.96/√T so significance occurs at ± 0.059 and these are imposed where appropriate.

**Table 2: Diagnostics for pre and post BEKK Modeling**

|  | Equity REITs | Mortgage REITs | Hybrid REITs | S&P 500 | S&P Value | S&P Growth | NASDAQ |
|---|---|---|---|---|---|---|---|
| Q(24)-R | 27.431 | 62.614 | 21.701 | 39.721 | 44.590 | 34.511 | 36.357 |
|  | (0.285) | (0.000) | (0.597) | (0.023) | (0.007) | (0.076) | (0.051) |
| $Q^2(24) - R$ | 388.713 | 254.260 | 138.854 | 196.644 | 313.115 | 142.974 | 376.347 |
|  | (0.000) | (0.000) | (0.000) | (0.000) | (0.000) | (0.000) | (0.000) |
| Q(24)-Z | 19.132 | 17.669 | 23.860 | 37.785 | 40.174 | 29.394 | 26.128 |
|  | (0.745) | (0.819) | (0.470) | (0.036) | (0.020) | (0.206) | (0.347) |
| $Q^2(24) - Z$ | 11.639 | 30.832 | 24.644 | 14.342 | 17.764 | 23.724 | 22.829 |
|  | (0.984) | (0.159) | (0.425) | (0.939) | (0.814) | (0.477) | (0.530) |
| ARCH(24) - R | 181.053 | 286.174 | 90.218 | 91.161 | 122.484 | 76.875 | 136.792 |
|  | (0.000) | (0.000) | (0.000) | (0.000) | (0.000) | (0.000) | (0.000) |
| ARCH(24) - Z | 11.357 | 32.543 | 23.032 | 15.278 | 19.185 | 23.471 | 24.279 |
|  | (0.986) | (0.114) | (0.518) | (0.912) | (0.742) | (0.492) | (0.446) |
| JB - R/Z | 1244.697 | 5270.718 | 106.178 | 23.030 | 39.667 | 20.664 | 5.909 |
|  | (0.000) | (0.000) | (0.000) | (0.000) | (0.000) | (0.000) | (0.052) |
| Skewness R/Z | -0.354 | -1.133 | -0.133 | 0.060 | 0.056 | 0.141 | 0.068 |
| Kurtosis – R/Z | 8.139 | 13.432 | 4.492 | 3.696 | 3.919 | 3.606 | 3.330 |

Notes: Q(24) – R and Q(24) – Z are Ljung-Box tests on the returns and residual series respectively whereas Q2(24) – R and Q2(24) – Z are Ljung-Box tests on the squared returns and residuals. ARCH(24) is the Engle (1981) LM test for up to twenty fourth order ARCH on the returns (R) and residual (Z) series. Skewness, kurtosis and the Jarque-Bera test for normality is shown for the standardized residual, R/Z, series. Marginal significance levels displayed in parenthesis.

## Table 3: Bivariate daily GARCH parameter estimates for REIT returns

| | Equity REITs-Mortgage REITs | | Equity REITs-Hybrid REITs | | Mortgage REITs-Hybrid REITs | |
|---|---|---|---|---|---|---|
| **Panel A: Conditional Mean** | | | | | | |
| $c_{11}$ | 0.000 | (0.405) | 0.000 | (0.426) | 0.000 | (0.272) |
| $r_{11}$ | 0.167 | (5.295) | 0.161 | (4.616) | 0.074 | (2.367) |
| $r_{21}$ | 0.020 | (1.344) | 0.030 | (1.160) | 0.081 | (1.722) |
| $c_{22}$ | 0.000 | (0.278) | 0.000 | (0.269) | 0.000 | (0.270) |
| $r_{22}$ | 0.072 | (2.303) | 0.070 | (2.017) | 0.113 | (3.640) |
| $r_{12}$ | 0.124 | (1.895) | 0.180 | (3.791) | 0.046 | (2.257) |
| **Panel B: Conditional variance** | | | | | | |
| $c_{11}$ | 0.002 | (6.221) | 0.002 | (8.455) | 0.001 | (1.892) |
| $c_{12}$ | 0.001 | (2.173) | 0.002 | (6.602) | 0.002 | (2.698) |
| $c_{22}$ | 0.002 | (5.328) | 0.003 | (17.249) | 0.003 | (13.888) |
| $a_{11}$ | 0.433 | (11.420) | 0.433 | (11.968) | 0.395 | (9.871) |
| $a_{12}$ | 0.016 | (1.189) | 0.051 | (1.556) | 0.021 | (0.409) |
| $a_{21}$ | 0.102 | (0.665) | 0.251 | (4.813) | 0.052 | (1.374) |
| $a_{22}$ | 0.358 | (7.613) | 0.251 | (6.741) | 0.235 | (10.286) |
| $b_{11}$ | 0.827 | (42.712) | 0.807 | (72.051) | 0.923 | (318.064) |
| $b_{12}$ | 0.000 | (0.059) | -0.030 | (-5.058) | -0.048 | (-1.711) |
| $b_{21}$ | -0.089 | (-0.894) | -0.132 | (-5.596) | -0.015 | (-1.150) |
| $b_{22}$ | 0.936 | (457.765) | 0.923 | (592.390) | 0.931 | (179.246) |

Notes: Quasi-maximum likelihood estimation gives robust t-statistics for the BEKK model based on Bollerslev and Wooldridge (1992) standard errors. T-statistics for the model are given in (). Own (other) market effects are given by 11 (22). Cross market effects to (from) own series are given by 21 (12).

**Table 4: Bivariate Daily GARCH Parameter Estimates Between REIT and Equity returns**

**Panel A: Equity REITs**

| | S&P 500 | | S&P Value | | S&P Growth | | NASDAQ | |
|---|---|---|---|---|---|---|---|---|
| | \multicolumn{8}{c}{Conditional Mean} | | | | | | | |
| $C_{11}$ | 0.000 | (-0.509) | 0.000 | (-0.353) | 0.000 | (-0.365) | 0.000 | (-0.360) |
| $R_{11}$ | -0.040 | (-1.155) | -0.025 | (-0.707) | -0.039 | (-1.192) | 0.006 | (0.198) |
| $R_{21}$ | 0.122 | (1.851) | 0.161 | (2.436) | 0.073 | (1.018) | 0.010 | (0.091) |
| $C_{22}$ | 0.000 | (0.378) | 0.000 | (0.402) | 0.000 | (0.384) | 0.000 | (0.352) |
| $R_{22}$ | 0.216 | (6.222) | 0.220 | (6.139) | 0.208 | (6.287) | 0.206 | (6.318) |
| $R_{12}$ | -0.035 | (-1.941) | -0.036 | (-1.877) | -0.030 | (-1.927) | -0.022 | (-2.220) |
| | \multicolumn{8}{c}{Conditional Variance} | | | | | | | |
| $C_{11}$ | 0.002 | (8.951) | 0.001 | (0.478) | 0.002 | (2.479) | 0.002 | (6.232) |
| $C_{12}$ | 0.001 | (1.516) | 0.001 | (0.814) | 0.002 | (1.245) | 0.002 | (2.648) |
| $C_{22}$ | 0.002 | (7.826) | 0.003 | (1.302) | 0.002 | (9.780) | 0.003 | (9.790) |
| $A_{11}$ | 0.213 | (8.775) | 0.216 | (0.905) | 0.226 | (6.176) | 0.215 | (12.248) |
| $A_{12}$ | 0.177 | (2.408) | 0.220 | (0.498) | 0.135 | (1.497) | 0.213 | (1.869) |
| $A_{21}$ | 0.016 | (0.391) | 0.026 | (0.197) | -0.001 | (-0.024) | -0.011 | (-0.741) |
| $A_{22}$ | 0.418 | (7.875) | 0.424 | (9.148) | 0.427 | (10.187) | 0.462 | (11.642) |
| $B_{11}$ | 0.973 | (3489.946) | 0.994 | (11424.80) | 0.967 | (924.492) | 0.976 | (2788.343) |
| $B_{12}$ | -0.096 | (-2.456) | -0.165 | (-3.899) | -0.084 | (-2.208) | -0.148 | (-2.811) |
| $B_{21}$ | 0.011 | (1.931) | 0.028 | (0.220) | 0.002 | (0.100) | 0.008 | (2.030) |
| $B_{22}$ | 0.830 | (79.612) | 0.779 | (5.060) | 0.847 | (93.208) | 0.815 | (79.648) |

**Panel B: Mortgage REITs**

| | S&P 500 | | S&P Value | | S&P Growth | | NASDAQ | |
|---|---|---|---|---|---|---|---|---|
| | \multicolumn{8}{c}{Conditional Mean} | | | | | | | |
| $c_{11}$ | 0.000 | (-0.547) | 0.000 | (-0.377) | 0.000 | (-0.408) | 0.000 | (-0.419) |
| $r_{11}$ | -0.016 | (-0.501) | 0.011 | (0.363) | -0.028 | (-0.908) | 0.009 | (0.308) |
| $r_{21}$ | 0.017 | (0.595) | 0.022 | (0.801) | 0.008 | (0.262) | -0.007 | (-0.138) |
| $c_{22}$ | 0.000 | (0.323) | 0.000 | (0.310) | 0.000 | (0.313) | 0.000 | (0.290) |
| $r_{22}$ | 0.075 | (2.433) | 0.077 | (2.486) | 0.077 | (2.498) | 0.084 | (2.761) |
| $r_{12}$ | 0.056 | (1.642) | 0.044 | (1.240) | 0.053 | (1.785) | 0.019 | (0.996) |
| | \multicolumn{8}{c}{Conditional Variance} | | | | | | | |
| $c_{11}$ | 0.001 | (0.460) | 0.000 | (0.825) | 0.003 | (3.787) | 0.002 | (5.468) |
| $c_{12}$ | -0.002 | (-1.458) | -0.002 | (-4.247) | -0.001 | (-0.116) | 0.001 | (3.125) |
| $c_{22}$ | 0.001 | (1.803) | 0.001 | (2.571) | 0.001 | (2.029) | 0.002 | (4.878) |
| $a_{11}$ | 0.229 | (2.996) | 0.235 | (6.774) | 0.247 | (9.655) | 0.225 | (18.067) |
| $a_{12}$ | -0.001 | (-0.022) | 0.008 | (0.189) | -0.026 | (-0.485) | 0.012 | (0.212) |
| $a_{21}$ | -0.145 | (-5.127) | -0.169 | (-6.146) | -0.094 | (-2.177) | -0.018 | (-1.291) |
| $a_{22}$ | 0.476 | (9.015) | 0.507 | (9.511) | 0.450 | (8.322) | 0.403 | (8.548) |
| $b_{11}$ | 0.954 | (794.283) | 0.955 | (1080.531) | 0.950 | (361.706) | 0.969 | (19293.57) |
| $b_{12}$ | 0.009 | (0.510) | 0.005 | (0.444) | 0.014 | (0.936) | -0.003 | (-0.207) |
| $b_{21}$ | 0.051 | (4.853) | 0.056 | (6.348) | 0.029 | (0.762) | 0.002 | (0.599) |
| $b_{22}$ | 0.892 | (130.022) | 0.882 | (114.549) | 0.907 | (118.106) | 0.925 | (330.497) |

**Panel C: Hybrid REITs**

| | S&P 500 | | S&P Value | | S&P Growth | | NASDAQ | |
|---|---|---|---|---|---|---|---|---|
| | \multicolumn{8}{c}{Conditional Mean} | | | | | | | |
| $c_{11}$ | 0.000 | (-0.502) | 0.000 | (-0.340) | 0.000 | (-0.368) | 0.000 | (-0.372) |
| $r_{11}$ | -0.002 | (-0.055) | 0.019 | (0.568) | -0.012 | (-0.363) | 0.025 | (0.797) |
| $r_{21}$ | -0.037 | (-0.812) | 0.000 | (0.001) | -0.078 | (-1.573) | -0.164 | (-2.222) |
| $c_{22}$ | 0.000 | (0.276) | 0.000 | (0.298) | 0.000 | (0.265) | 0.000 | (0.253) |
| $r_{22}$ | 0.127 | (3.910) | 0.115 | (3.520) | 0.135 | (4.256) | 0.135 | (4.327) |
| $r_{12}$ | 0.014 | (0.596) | 0.036 | (1.467) | -0.004 | (-0.208) | -0.003 | (-0.244) |
| | \multicolumn{8}{c}{Conditional Variance} | | | | | | | |
| $c_{11}$ | 0.002 | (7.374) | 0.002 | (0.185) | 0.002 | (5.719) | 0.002 | (1.749) |
| $c_{12}$ | 0.002 | (2.897) | 0.000 | (-0.013) | 0.002 | (4.591) | 0.002 | (0.703) |
| $c_{22}$ | 0.003 | (13.843) | 0.005 | (0.641) | 0.003 | (12.417) | 0.003 | (4.021) |
| $a_{11}$ | 0.247 | (9.763) | 0.248 | (0.217) | 0.239 | (8.828) | 0.236 | (11.927) |
| $a_{12}$ | 0.044 | (0.502) | -0.009 | (-0.009) | 0.052 | (0.414) | 0.071 | (0.079) |
| $a_{21}$ | 0.058 | (1.394) | 0.110 | (0.480) | 0.034 | (0.865) | 0.029 | (0.586) |
| $a_{22}$ | 0.307 | (6.527) | 0.343 | (5.976) | 0.318 | (7.261) | 0.313 | (2.389) |
| $b_{11}$ | 0.956 | (1727.000) | 0.936 | (7.665) | 0.963 | (1486.479) | 0.970 | (1611.987) |
| $b_{12}$ | -0.041 | (-0.868) | 0.055 | (0.029) | -0.067 | (-0.839) | -0.085 | (-0.178) |
| $b_{21}$ | -0.012 | (-1.003) | 0.037 | (0.038) | -0.009 | (-0.535) | -0.003 | (-0.209) |
| $b_{22}$ | 0.891 | (115.991) | 0.709 | (0.617) | 0.888 | (62.523) | 0.886 | (14.977) |

Notes: Quasi-maximum likelihood estimation gives robust t-statistics for the BEKK model based on Bollerslev and Wooldridge (1992) standard errors. T-statistics for the model are given in (). Own (other) market effects are given by 11 (22). Cross market effects to (from) own series are given by 21 (12).

## Table 5: Summary statistics of daily REIT correlation

|  | Equity REITs- Mortgage REITs | Equity REITs- Hybrid REITs | Mortgage REITs- Hybrid REITs |
|---|---|---|---|
| **Min** | -0.512 | -0.005 | -0.140 |
| **1st quartile** | 0.167 | 0.314 | 0.134 |
| **Median** | 0.279 | 0.411 | 0.265 |
| **3rd quartile** | 0.411 | 0.535 | 0.404 |
| **Max** | 0.909 | 0.919 | 0.865 |
| **Mean** | 0.294 | 0.436 | 0.279 |
| **Std Deviation** | 0.208 | 0.170 | 0.198 |
| **Skewness** | 0.287 | 0.542 | 0.423 |
| **Kurtosis** | 0.453 | 0.049 | -0.214 |
| **Normality** | 0.061 | 0.063 | 0.039 |

Note: Table 5 gives the summary statistics for daily REIT correlations. Correlations are obtained from fitting the BEKK specification. It also presents the results of the Kolmogorov-Smirnov test for normality. All skewness, kurtosis and normality test coefficients are significant.

**Figure 4: Time Series Plots for Daily Conditional Correlations for REIT Returns**

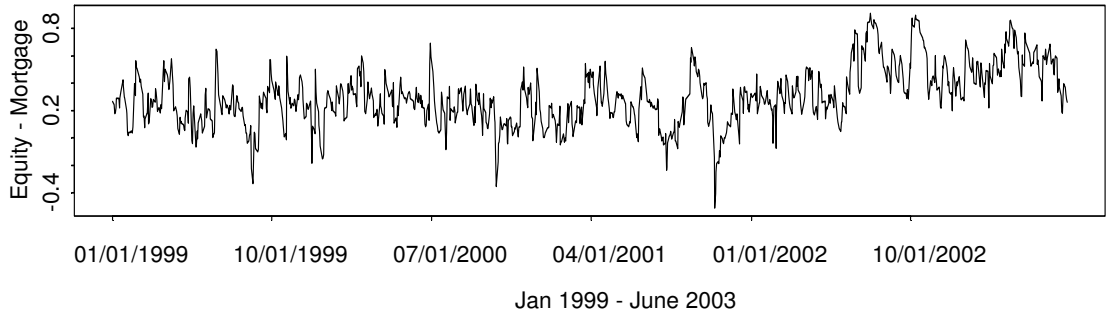

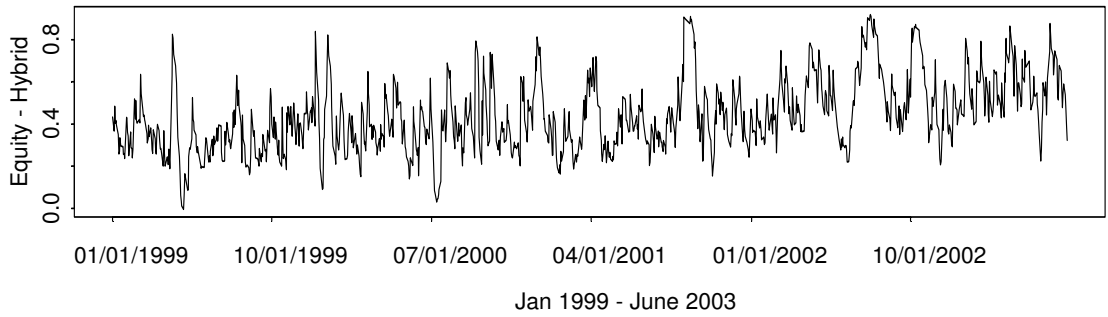

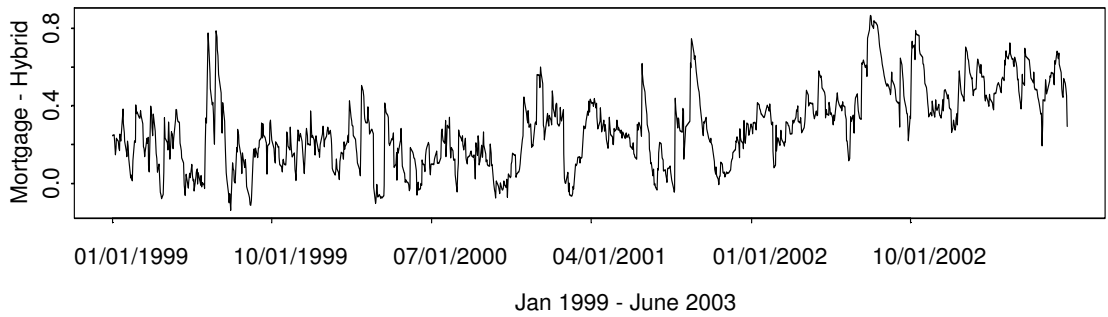

**Figure 5: Time Series Plots for Daily Conditional Correlations for S&P500 and REIT Returns**

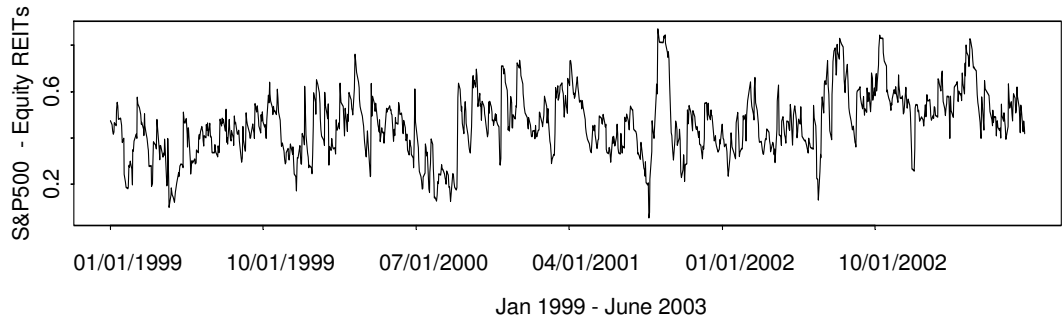

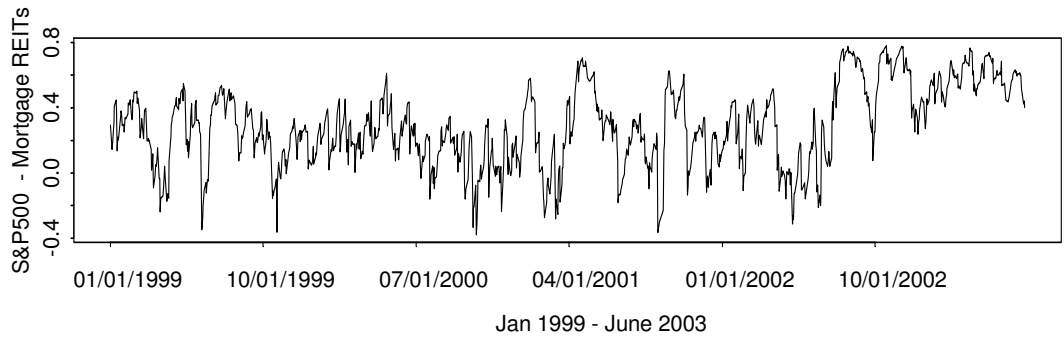

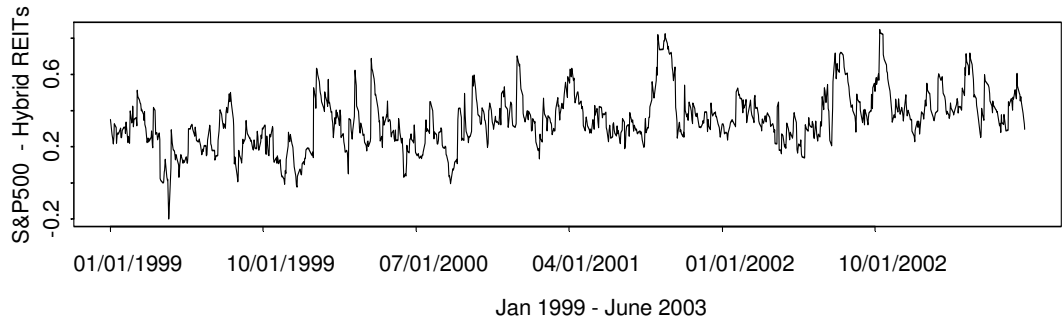

**Endnotes:**

[1] See for example, Barkham & Geltner (1995).

[2] The authors own calculations.

[3] The ADF results are not reported for conciseness but are available from the authors.

[4] The other equity series follow a similar pattern to the Equity and Hybrid REIT series' and the corresponding figures are available from the authors. The range of returns is reasonably similar between the REIT and S&P series' although the small capitalization companies captured by the S&P Value index are capable of larger absolute returns. The NASDAQ also is more volatile than the REIT series, however, given the time period examined this not surprising. The sample period includes both the final 15 months of the technology bubble and the following crash. Regime change may have occurred as a result of the technology crash. Using a dummy for the April 2000 technology crash we find a significant effect for the Equity REIT sector but not for the others. The authors are currently investigating this issue further using regime-switching models on a longer time frame with monthly returns.

[5] Other multivariate GARCH models do not have the relative advantages of the BEKK process in directly isolating the bivariate relationships. However these models do allow for a comparison with the BEKK approach and this is currently being explored in another paper by the authors. A distinct alternative to the multivariate GARCH approach would be for example the development of a stochastic volatility model. Bond & Hwang (2003) use a stochastic volatility approach in their analysis of the fundamental volatility of the commercial real estate market in the UK.

[6] While the multivariate model selected would allow a broader analysis than a simple bivariate one, diagnostic issues did arise due to the similarity between many of the equity indices involved.

[7] The corresponding conditional correlations with the other equity series' are available from the authors. These are broadly consistent with the findings for the S&P 500.

[8] Ling & Naranjo (2004) illustrate the impact of the flow of funds into REITs on their returns.